\documentclass{article} 
\usepackage[margin=1in]{geometry}
\usepackage{authblk}

\usepackage{natbib}
\usepackage[utf8]{inputenc} 
\usepackage{hyperref}
\usepackage{url}
\usepackage{amsfonts}       
\usepackage{amstext}
\usepackage{amsmath,amsthm}
\usepackage{xcolor}         
\usepackage[linesnumbered,ruled]{algorithm2e}   
\usepackage{multirow}
\usepackage{graphicx}
\usepackage{comment}
\usepackage{tablefootnote}

\usepackage{hyperref} 
\usepackage{lineno}

\usepackage{floatrow}
\usepackage{wrapfig}

\usepackage{hyperref}       
\usepackage{url}            
\usepackage{booktabs}       
\usepackage{amsfonts}       
\usepackage{nicefrac}       
\usepackage{microtype}      
\usepackage{xcolor}         
\usepackage{bm}             
\usepackage{mathrsfs}
\usepackage{xr}
\usepackage[T1]{fontenc}

\title{Survey on Recent Progress of {\em AI for Chemistry}:\\
Methods, Applications, and Opportunities}

\date{}

\begin{document}

\author[1]{Hu Ding}
\author[2]{Pengxiang Hua}
\author[2]{Zhen Huang\thanks{Authors are listed in alphabetical order by last name.}}

\affil[1]{University of Science and Technology of China, Hefei, China 

(\texttt{huding@ustc.edu.cn})}

\affil[2]{University of Science and Technology of China, Hefei, China 

(\texttt{\{hpx, huangzhen\}@mail.ustc.edu.cn})}

\maketitle

\begin{abstract}
The development of artificial intelligence (AI) techniques has brought revolutionary changes across various realms. In particular, the use of AI-assisted methods to accelerate chemical research has become a popular and rapidly growing trend, leading to numerous groundbreaking works. In this paper, we provide a comprehensive review of current AI techniques in chemistry from a computational perspective, considering various aspects in the design of methods. We begin by discussing the characteristics of data from diverse sources, followed by an overview of various representation methods. Next, we review existing models for several topical tasks in the field, and conclude by highlighting some key challenges that warrant further attention. 
\end{abstract}

\section{Introduction}

\emph{Artificial Intelligence} (AI) refers to a series of techniques that can perform tasks typically requiring human intelligence, such as speech recognition, language translation, path planning, and so on. This concept originated at the 1956 Dartmouth conference, and over the decades, AI technologies have become pervasive in a variety of areas and achieved remarkable results. In addition to applications in commerce or industry, people have been exploring how AI can be utilized to accelerate scientific research and discovery~\cite{stevens_ai_2020}. For example, the AlphaFold developed by DeepMind can accurately predict the three-dimensional structure of proteins based on amino acid sequence, revolutionizing the field of structural biology~\cite{jumper_highly_2021}. George et al. combined neural networks with physical laws expressed as PDEs, enabling efficient solutions to complex high-dimensional problems in fluid dynamics, quantum mechanics, and more~\cite{raissi2019physics}. AI for Science has become a new paradigm in scientific research, and has the potential to surpass the previous heuristics-based experimental design and discovery.

{\em Chemistry} is the fundamental science that drives technological innovation by unlocking the secrets of matter and enabling the discovery of groundbreaking materials~\cite{garcia2011chemical}. Traditional chemistry research often faces significant challenges, such as the time-consuming trial-and-error processes, the complexity of modeling molecular interactions, and the vast number of possible reaction combinations. To accelerate the process of chemical research, there has been a growing exploration of the potential applications of AI in the field of chemistry. Developed at Stanford University in the mid-1960s, ``DENDRAL'' was one of the earliest and most influential chemical expert systems~\cite{buchanan1969heuristic}. By encoding the knowledge of experienced chemists into a series of rules and heuristics, the system was successfully applied to identify molecular structures from mass spectrometry data. Subsequently, several other expert systems emerged for predicting the reaction outcomes~\cite{salatin1980computer, funatsu1988computer, satoh1995sophia}, but all relied on predefined rules and were limited in adapting to new or complex scenarios.

Recent advances in computational power and the development of automated synthesis technologies have led to a growing interest in developing \emph{machine learning (ML)} methods for chemistry. For example, Ahneman et al. conducted around 4600 ``Pd-catalyzed Buchwald-Hartwig'' reactions using high-throughput equipment, and validated the yield prediction performance of several ML models such as linear regression and random forest~\cite{ahneman2018predicting}. One of the most fascinating techniques in machine learning today is {\em large language models (LLMs)}~\cite{brown2020language, achiam2023gpt, touvron2023llama, zhao2023survey}, and meanwhile LLMs for chemistry is gradually gaining attention. For instance, several LLMs enabled agent systems for chemical research have been proposed~\cite{m2024augmenting, boiko2023autonomous}. The powerful encoding and feature extraction abilities of language models are also utilized to improve downstream tasks~\cite{schwaller2021prediction, schwaller2020data}.

\begin{figure}[!h]
    \centering
    \includegraphics[width=\textwidth]{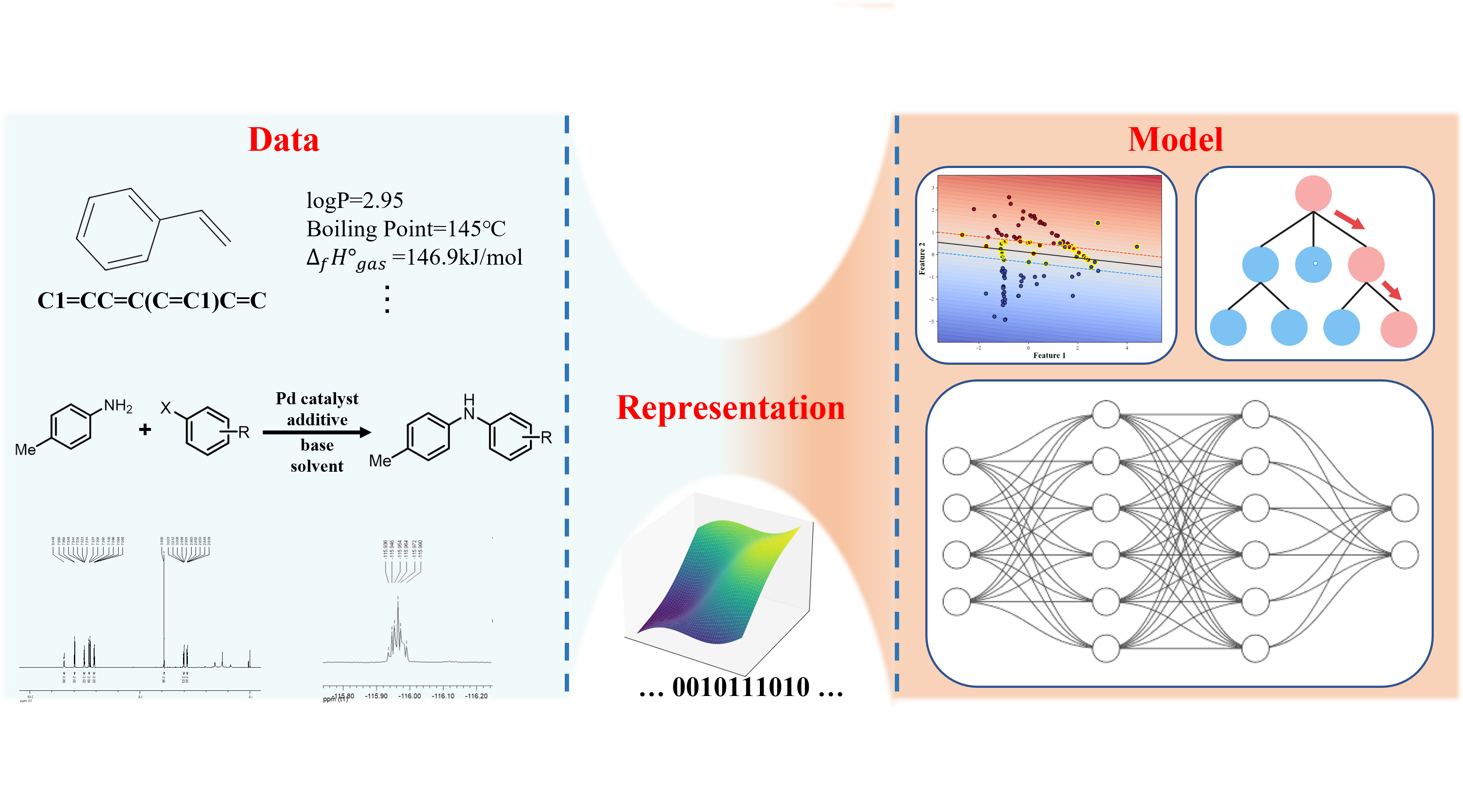}
    \caption{\textbf{Three components in machine learning: data, representation and model.}}
    \label{fig:data_rep_model}
\end{figure}

To provide a thorough introduction to the advancement of \emph{AI for Chemistry} from the lens of computer science, it is essential to first clarify the paradigm of AI research, particularly in machine learning. The goal of ML is to extract knowledge from data to assist machines in performing certain tasks. Generally, ML involves three important components: \textbf{data}, \textbf{representation} and \textbf{model} (Fig.~\ref{fig:data_rep_model}). Data serves as the foundation for ML, enabling the identification of patterns and relationships. For evaluating the accuracy and generalization of ML models, high-quality and diverse data is crucial. The collected raw data should be transformed into an appropriate machine-recognizable format, referred to as \emph{representation}. An effective representation captures the essential features of the data while filtering out irrelevant or noisy information. The model, which is typically a mathematical framework equipped with construction algorithms, learns patterns from the training data and applies them to make predictions or decisions in new scenarios. Models can range from simple approaches, such as polynomial or tree-based models, to complex neural networks~\cite{liu2017survey, pouyanfar2018survey} and large language models. Moreover, how to select an appropriate model also heavily depends on the task at hand.

\textbf{The remainder of this article is organized as follows:} First, we introduce various data sources in the field of chemistry and discuss their characteristics. Next, we present the commonly used representation methods in machine learning. We then summarize the existing approaches for key application scenarios, including prediction tasks, molecular design, retrosynthesis, and AI-driven chemical robots. Following that, we explore recent studies on LLMs in chemistry. Finally, to support future advancements in this field, we highlight several key challenges in the application of AI techniques to chemistry.

\section{{Datasets} and Descriptions}

\begin{figure}[t]
    \centering
    \includegraphics[width=0.8\textwidth]{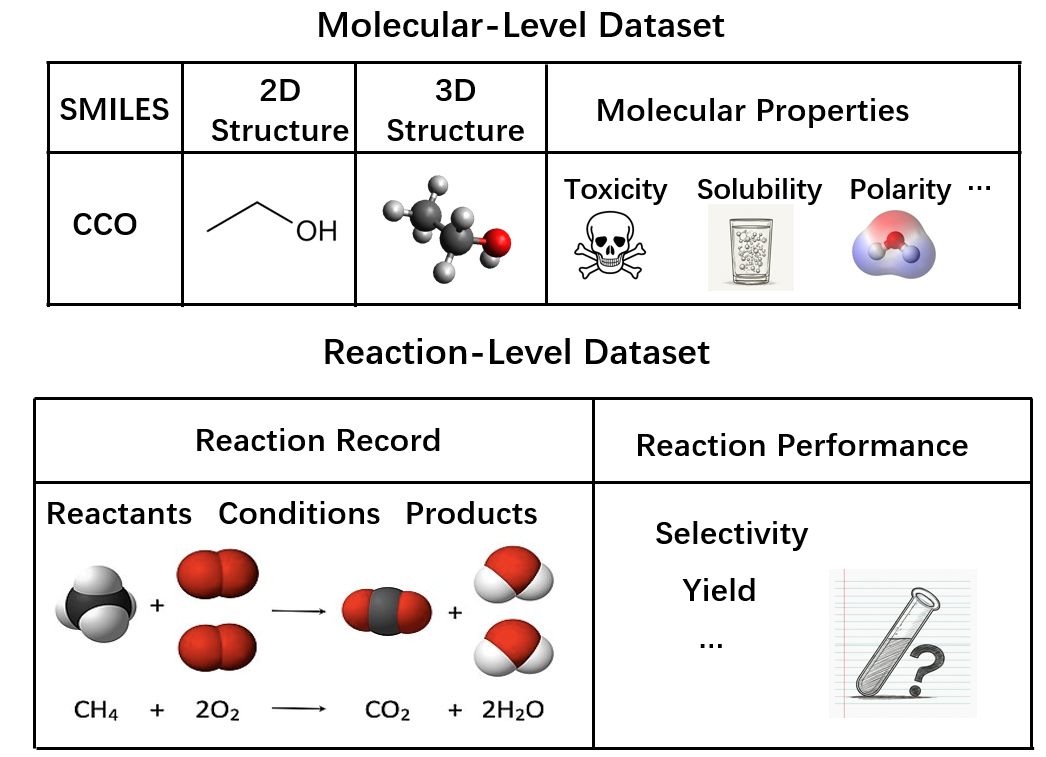}
    \caption{\textbf{Different types of chemical datasets.}}
    \label{fig:Datasets}
\end{figure}

Machine learning is a data-driven discipline, and its rapid progress has been largely propelled by the increasing availability of data~\cite{zhou2017machine}. 
In the field of chemistry, addressing various data challenges is particularly demanding~\cite{strieth2022machine}.
High-quality, sufficiently large datasets are essential for machine learning techniques to uncover and understand chemical principles. 
However, as an experimental science, chemistry often relies on data derived from laboratory experiments or computational simulations, which can result in issues such as data scarcity and inconsistent quality~\cite{feller1996role}.
Moreover, data are typically recorded in the individual laboratory, making it difficult to establish large-scale, publicly available datasets~\cite{liu2023data}.  
Fortunately, significant advancements in technology and research over the years have led to substantial improvements in chemical datasets, providing a great support for machine learning applications in this field.
Chemical datasets generally fall into two categories: \textbf{molecular-level} datasets, which record molecular properties and features, and \textbf{reaction-level} datasets, which focus on reaction conditions and outcomes (Fig.~\ref{fig:Datasets}).
This section will introduce several widely used datasets in the field, and a summary of these datasets is shown in Table~\ref{tab:dataset}.

\subsection{Descriptions}

Chemical datasets contain rich information, including molecular properties and reaction yields. 
However, to fully capture and utilize this information, we need first identify molecular entities and translate abstract molecular concepts into machine-readable formats.
Therefore, before introducing the datasets, we will first discuss some common  molecular identifiers and descriptors.

\subsubsection{Molecular Identifiers}
{\em Molecular identifiers} are symbols, strings, or codes that uniquely represent chemical molecules.
They enable the systematically representation, storage, sharing, and retrieval of chemical substances.
Traditionally, molecules are identified using identifiers such as the \textit{Chemical Abstracts Service (CAS)} registry Number or database retrieval IDs.
In the context of machine learning, however, a chemical identifier not only serves as a unique label for a molecule but also encodes basic chemical information.
There are several common examples of these molecular identifiers,  such as the {\em Simplified Molecular Input Line Entry System (\textbf{SMILES})}~\cite{weininger1988smiles} and the {\em International  Chemical Identifier (\textbf{InChI})}~\cite{heller_inchi_2015}. 
These identifiers describe a molecule’s structure, composition, and other key features, and are widely used in machine learning models as inputs for various tasks.

\textbf{SMILES}

\begin{figure}[t]
    \centering
    \includegraphics[width=0.4\textwidth]{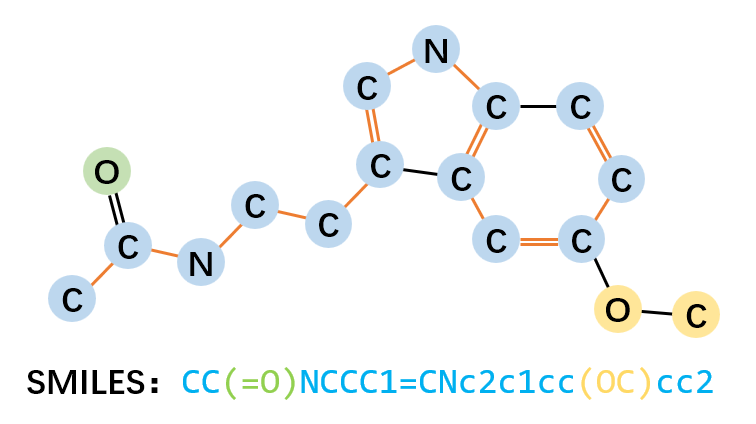}
    \caption{\textbf{An example of the mapping between SMILES and structure.}
    The molecule in the figure is ``melatonin''. Hydrogen atoms are omitted, and atoms in different colors correspond to the text in the SMILES notation, representing the main chain and side chains, respectively. The orange bonds highlight the edges of the main chain split by SMILES.}
    \label{fig:Smiles}
\end{figure}

SMILES is a notation system designed to represent the structure of  molecules in a human-readable text format (Fig.~\ref{fig:Smiles}). 
Its primary goal is to offer a simple way to encode molecular structures while ensuring compatibility with computational tools. 
In SMILES, atomic symbols represent elements, with bonds typically implied or denoted by specific symbols such as '=', '\#'.
This system efficiently captures branching and ring structures and can encode stereochemistry and isotopes when necessary. 

As a human-readable identify, SMILES is ideal for molecular data storage and retrieval; due to its compact nature and ease of use, it has been widely adopted in cheminformatics and computational chemistry.
However, SMILES can encounter challenges when representing certain complex molecules. Additionally, it lacks a standard approach for representing stereochemical information, and in some cases, SMILES identifiers may not be unique. To address these limitations, alternative or improved linear representation methods, such as \textit{SMILES Arbitrary Target Specification (SMART)}~\cite{daylight_smarts} and \textit{Self Referencing Embedded String (SELFIES)}~\cite{krenn2020self}, have been proposed.

\textbf{InChI}

InChI is a standardized molecular identifier developed by the \textit{International Union of Pure and Applied Chemistry (IUPAC)} to uniquely represent chemical substances in a machine-readable format.
It encodes the structural information of a molecule in a layered format, detailing aspects such as atomic connectivity, stereochemistry, isotopic composition, and charge states~\cite{heller_inchi_2015}.
The InChI format ensures that each chemical substance is represented uniquely, making it an essential tool for chemical databases and research.
Its machine-readable nature facilitates seamless data exchange across different systems, enhancing consistency and accessibility.

InChI has become essential in global chemical research, particularly in drug discovery, as it standardizes data formats across various databases and platforms.
However, it has limitations in representing complex structures compared to other identifiers like SMILES, and it is also more difficult for humans to interpret.
The InChI standard is still under continuous development by IUPAC and the InChI Trust, with ongoing efforts to enhance and expand its applicability.

\subsubsection{Molecular Fingerprints}

{\em Molecular fingerprints} are widely used in cheminformatics to enable the comparison and classification of molecules based on their structural features.
These fingerprints typically represent molecular information as binary or integer vectors, where each bit or value indicates the presence or absence of specific structural features, such as functional groups, atoms, or bonds.
Different types of molecular fingerprints exist, each capturing distinct aspects of a molecule, from basic atomic features to more complex topological and stereochemical information.
The following are some common types of molecular fingerprints.

{\em AtomPair Fingerprints} capture the relationships between pairs of atoms in a molecule, considering both atom types and their distances~\cite{carhart1985atom}. 
The \textit{Molecular ACCess System (MACCS)} keys, developed by Durant et al., consist of a set of predefined substructures (typically 166), encoding the presence or absence of each substructure within a molecule \cite{durant2002reoptimization}. 
{\em Pharmacophore Fingerprints}, proposed by Wpolber et al., focus on the spatial arrangement of functional groups or pharmacophores, which are key factors in determining a molecule's biological activity~\cite{wolber2008molecule}. \textit{Extended-Connectivity Fingerprints (ECFP)}, also known as {\em Morgan} fingerprints, are circular fingerprint generated by iteratively exploring the environment around each atom \cite{rogers2010extended}.
ECFPs are highly customizable in terms of radius and bit size, which allows them to capture detailed molecular information.
These types of fingerprints are widely applied in similarity searching~\cite{cereto2015molecular}, virtual screening~\cite{melville2009machine},  drug discovery and molecular property prediction~\cite{sandfort2020structure, doi:10.1021/acs.accounts.0c00699}.

\subsection{Molecular-Level Datasets}

Molecular-level datasets are fundamental to the application of machine learning in fields like chemistry, drug discovery, and materials science. 
These datasets typically include molecular structures, quantum mechanical properties, physicochemical properties, biological activity data, and other relevant information. They are sourced from various outlets, including scientific publications, patents, or commercial databases. 
Such datasets can be used to evaluate a model's ability to predict molecular properties and can also serve as training data for developing molecular representations. 

\subsubsection{Several Common Public Databases}

Based on centuries of continuous exploration and discovery, we witness that several well-established chemical databases have been created and are now publicly accessible.
These datasets contain millions of molecular records, with the information of molecular structure, chemical formula, SMILES and InChI. 
Prominent examples include ChEMBL \cite{zdrazil2024chembl}, PubChem~\cite{kim2025pubchem} and ZINC~\cite{irwin2020zinc20}.
ChEMBL is one of the largest and most well-established bioactivity databases, offering a comprehensive resource to the scientific community.
It provides extensive data on the bioactivity of small molecules, focusing on their interactions with biological targets such as proteins, enzymes, and receptors.
ZINC specializes in providing researchers with a vast collection of commercially available compounds, ideal for virtual screening.
PubChem, maintained by the \textit{National Center for Biotechnology Information (NCBI)}, is a comprehensive chemical database that offers detailed information on the biological activities, structures, and properties of chemical compounds.

\subsubsection{Quantum Mechanics Datasets}
{\em Quantum Mechanics (QM)} is a fundamental theory in physics that describes the behavior of matter and energy at the atomic and subatomic levels~\cite{dirac1981principles}. 
In computational chemistry, QM methods are employed to predict the electronic structure of molecules and materials, providing valuable insights into their properties and behaviors. 
These methods have become indispensable tools for studying molecular interactions, chemical reactions, and material design~\cite{szabo1996modern,mcquarrie1997physical,jensen2017introduction}.
However, QM methods typically require substantial time and computational resources. For instance, performing a detailed electronic structure calculation using techniques such as Density Functional Theory (DFT) or coupled-cluster theory can take hours or even days on high-performance computing systems, especially for large molecules. 
As a result, using machine learning methods to efficiently and accurately predict quantum chemical properties holds significant practical value.

To support the development of these machine learning methods, several quantum chemistry datasets have been publicly released. {\em Quantum-machine}~\cite{quantum_machine} is an open project that provides a large collection of datasets with the goal of developing a ``quantum machine'' for fast and accurate quantum chemical simulations. 
The project has produced several popular quantum mechanics datasets, including QM7, QM7b, QM9, and QM8~\cite{blum,rupp,1367-2630-15-9-095003,ruddigkeit2012enumeration,ramakrishnan2014quantum,ramakrishnan2015electronic}. These datasets contain a weath of valuable molecular quantum chemical properties. 
For example, the QM9 dataset includes key properties such as energy, dipole moments, and free energies, making it a valuable resource for 
developing new computational techniques and exploring structure-property relationships in organic molecules.

\subsubsection{Molecular Dynamics Datasets}
{\em Molecular dynamics (MD)} is a computational simulation technique used to model the physical movements of atoms and molecules over time~\cite{alder_studies_1959}. 
By solving Newton's equations of motion, MD simulations provide valuable insights into the behavior of molecules at the atomic level, helping to understand their structures, interactions, and dynamics.
In machine learning, incorporating molecular dynamics data as input can enhance model performance in predicting related properties such as energy~\cite{doi:10.1021/acs.jcim.6b00778}. 
How to leverage ML techniques to accelerate MD predictions is an important issue that warrants attention, and there already exist several machine learning models that integrate with MD computational methods~\cite{doi:10.1126/sciadv.1603015, chmiela2018, chmiela2020}.
As an example, the MD17 dataset~\cite{doi:10.1126/sciadv.1603015} is a benchmark collection designed for training and evaluating machine learning models in molecular dynamics. It includes data for molecules such as benzene, toluene, naphthalene, ethanol, uracil, and aspirin~\cite{doi:10.1126/sciadv.1603015}. The dataset provides molecular dynamics trajectories generated via ab initio methods, which include atomic coordinates, potential energies, and atomic forces. 

\subsubsection{Molecular Properties Datasets}

Molecular properties, such as solubility, toxicity, and biological activity are also key factors in chemical research. 
For example, the Blood-Brain Barrier Penetration (BBBP) dataset~\cite{sakiyama_prediction_2021} is used to predict whether a molecule can cross the blood-brain barrier, which is a key factor in designing drugs that affects the central nervous system.
The FreeSolv dataset~\cite{mobley_freesolv_2014} provides solvation free energy data for molecules in water, offering valuable insights into how a molecule will interact in aqueous environment. 
This information aids in drug design by predicting the solubility of compounds.
The Tox21 dataset~\cite{Mayr2016,Huang2016} contains data on the toxicity of various compounds, specifically focusing on their effects on biological systems.

These molecular property datasets are widely studied and commonly used as benchmarks for evaluating molecular prediction models. However, since obtaining such data often requires conducting experimental works in chemical labs, these datasets are generally smaller in size compared to QM and MD datasets.

\begin{table}[!h]
    \centering
    \caption{Summary on several popular chemical datasets}
    \begin{tabular}{p{3cm}ccp{10cm}}
        \toprule
        Name & Type & Size & Description \\
        \midrule
        \textbf{ZINC}~\cite{irwin2020zinc20} & Molecule & 230M & A free database of over 230 million commercially-available compounds in ready-to-dock, 3D formats\\ 
        \textbf{MD17}~\cite{doi:10.1126/sciadv.1603015} & Molecule & 3M & Provides molecular dynamics trajectories generated via ab initio methods, including atomic coordinates, potential energies, and atomic forces. \\ 
        \textbf{QM7}~\cite{blum, rupp} & Molecule & 7K & A subset of GDB-13, provides the Coulomb matrix representation of molecules and their atomization energies computed. \\ 
        \textbf{QM9}~\cite{ruddigkeit2012enumeration, ramakrishnan2014quantum} & Molecule & 100K & Contains quantum chemical properties for 134,000 stable small organic molecules composed of CHONF, derived from a subset of the GDB-17 chemical universe. \\ 
        \textbf{QM8}~\cite{ruddigkeit2012enumeration, ramakrishnan2015electronic} & Molecule & 20K & Comprises low-lying singlet-singlet vertical electronic spectra for over 20,000 synthetically feasible small organic molecules with up to eight heavy atoms, predicted using time-dependent density functional theory (TDDFT). \\ 
        \textbf{Tox21}~\cite{Mayr2016,Huang2016} & Molecule & 7K & Tox21 dataset provides 12,060 training samples and 647 test samples, for each sample there are 12 binary labels that represent the outcome (active/inactive) of 12 different toxicological experiments. \\ 
        \textbf{BBBP}~\cite{sakiyama_prediction_2021} & Molecule & 2K & BBBP dataset contains binary labels indicating whether a compound can penetrate the blood-brain barrier or not. \\ 
        HIV~\cite{riesen2008iam} & Molecule & 40K & The HIV dataset was generated by Drug Therapeutics Program(DTP) AIDS Antiviral Screen, containing anti-HIV activity data of over 40,000 compounds. \\ 
        \textbf{MoleculeNet}~\cite{wu2018moleculenet} & Molecule & 1K-400K & A large scale benchmark for molecular machine learning, including multiple public datasets. \\ 
        \textbf{USPTO-50k}~\cite{marco2015uspto, schneider2016s} & Reaction & 50K & A subset and preprocessed version of chemical reactions from US patents (1976–Sep 2016), containing 50,000 randomly selected reactions that were later classified into 10 reaction classes. \\ 
        \textbf{USPTO-190}~\cite{marco2015uspto,chen2020retro} & Reaction & 300K & A chemical synthesis route dataset derived from the USPTO reaction dataset. \\
        \textbf{USPTO 1k TPL}~\cite{marco2015uspto,schwaller2021mapping} & Reaction & 445K & A reaction dataset sourced from the USPTO, consisting of 445k reactions categorized into 1000 template labels.\\
        \textbf{BH HTE}~\cite{ahneman2018predicting} & Reaction & 4K & This dataset records the reaction yields of thousands of BH reactions obtained using HTE technology. \\ 
        \textbf{SM HTE}~\cite{perera2018platform} & Reaction & 5K & This dataset records the reaction yields of thousands of SM reactions obtained using HTE technology. \\ 
        \bottomrule
    \end{tabular}
    \label{tab:dataset}
\end{table}

\subsection{Reaction-Level Datasets}

Reaction-level datasets typically document reaction conditions and components, sometimes also including selectivity or yield information.
In general, these datasets can be  constructed from three sources: existing literature and patents, laboratory experimental records (i.e. \textit{electronic laboratory notebooks (ELNs)}), and specifically designed \textit{high-throughput experiments (HTE)}~\cite{saebi2023use}.

To describe reaction-level datasets, we must first introduce the concept of ``\textbf{reaction space}'', which is fundamental in chemistry.
Reaction space refers to the multidimensional space that encompasses all possible reaction combinations, including reactants, products and reaction conditions. 
The complexity and diversity of reaction space often lead to varying degrees of data insufficiency~\cite{stocker2020machine, williams2021evolution}. 

\subsubsection{USPTO}

The USPTO reaction dataset contains millions of reactions extracted through text mining from United States patents published between 1976 and September 2016~\cite{marco2015uspto}. 
Each entry records a reaction in formats such as \textit{Chemical Markup Language (CML)} or SMILES, including the details for reactants, reagents, and conditions. 
Although the dataset contains over one million reactions, it suffers from issues such as data duplication, poor quality, and inconsistencies.
As a result, data cleaning is often necessary before using it for training purposes.
Additionally, the USPTO dataset has been applied in the development of retrosynthesis models. It also provides valuable insights into reaction conditions, catalysts, and reagents, thereby contributing to more efficient chemical processes and advancing computational chemistry~\cite{chen2020retro}.

\subsubsection{Reaxys}

Reaxys is a commercial chemical database that provides extensive reaction data sourced from patents and journals~\cite{reaxys, doi:10.1021/ci900437n}. 
It offers detailed information on reaction conditions and product yields, which is of great importance for synthetic planning and reaction optimization. 
Reaxys plays a crucial role in both academic and industrial research, assisting in the design of synthetic pathways, optimization of processes, and discovery of new reactions. 
The database is regularly updated and has supported a variety of applications
from reaction prediction to material and drug discovery~\cite{reaxys}.

\subsubsection{HTE Datasets}

{\em High-Throughput Experimentation (HTE)} accelerates the exploration and optimization of chemical processes through automation, reducing both the time and resource usage~\cite{shevlin2017practical, krska2017evolution}. 
Several HTE datasets, such as the Suzuki coupling dataset~\cite{perera2018platform} and Buchwald-Hartwig dataset~\cite{ahneman2018predicting}, provide great amounts of reaction data under varying conditions, supporting tasks like yield prediction and reaction optimization. 
These reaction datasets typically account for multiple variables in the reactions, generating a reaction space with thousands of data points.

\subsection{Additional Remarks on Chemical Data}
In general, the lower the cost of acquiring data, the larger the available dataset. 
Low-cost data often consists of simple, basic descriptions of molecules and is typically used for representation learning. 
In contrast, high-quality data, which is harder to obtain, captures more macroscopic and direct properties. Such data is highly valuable for prediction tasks in chemistry, as models generally require a sufficient amount of high-value data for effective training~\cite{strieth2022machine}.

To address data scarcity, machine learning methods like transfer learning~\cite{shim2022predicting} and active learning~\cite{shields2021bayesian, gong2021deepreac+} can be applied.
Transfer learning involves transferring knowledge from a pre-trained model on one task to a related task and is particularly useful when  the new task only has limited training data~\cite{pan2009survey}. 
Active learning is a machine learning technique that iteratively identifies  the most informative data points from an unlabeled dataset for labeling~\cite{ren2021survey}. 
Instead of randomly selecting data points, the model queries for labels on the most ``uncertain'' ones, and therefore yields a more efficient learning process.
Moreover, acquiring large volumes of high-quality data through HTE~\cite{shevlin2017practical, krska2017evolution} or leveraging ELN data~\cite{saebi2023use}
could also serve as effective strategies to mitigate the issue of data scarcity.
Nonetheless, solving the challenges surrounding data for chemistry remains one of the most critical issues today.

\section{Representation Learning Methods for Chemical Objects}

Effectively representing data is a critical challenge in machine learning~\cite{bengio2013representation}. 
In the field of chemistry, this challenge is even more pronounced due to the complexity and diversity of chemical properties. 
A well-designed representation should be capable of distinguishing different molecular properties, meaning the distribution in the representation space should strongly correlate with the downstream task~\cite{talanquer2022complexity}.
This requires that training data should contain sufficient chemical information, such as the topological relationships between atoms,  3D conformation and electronic distribution~\cite{kearnes2016molecular}.

In the previous section, we discussed simple molecular representations such as molecular fingerprints.
While these methods are effective and align well with the intuition of chemists, they often fall short in tasks that demand higher generalizability in representation. 
In the domain of AI for chemistry, \textbf{graph neural networks (GNNs)}\cite{gilmer2017neural,zhou2020graph,jiang2021could} are widely adopted as the foundational model for representation learning, given the natural relevance of molecular graphs to this architecture. Additionally, the \textbf{Transformer} architecture~\cite{vaswani2017attention}, which has shown significant promise in AI, is also increasingly used to process SMILES strings to obtain better molecular representations~\cite{raghunathan2022molecular}. Below, we delve into these two approaches in greater details. 

\subsection{GNN Based Representation Learning}

\begin{figure}[t]
    \centering
    \includegraphics[width=0.5\textwidth]{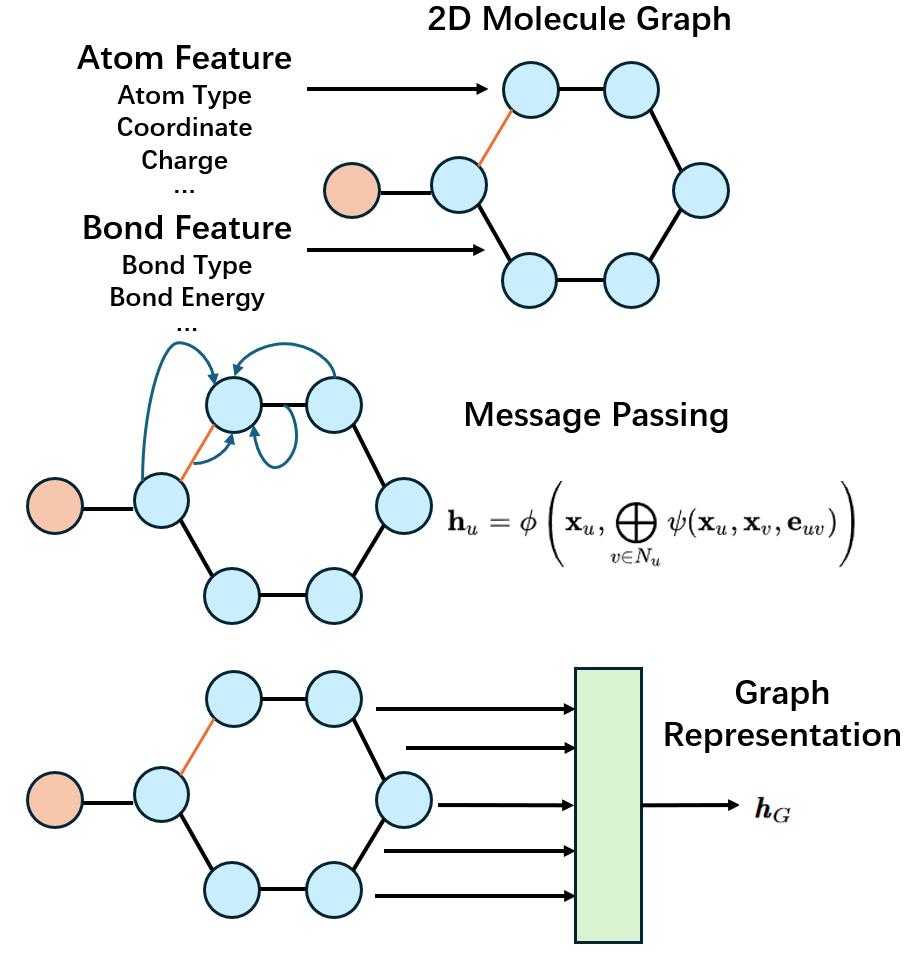}
    \caption {\textbf{A common framework of molecular graph neural networks.} Through message-passing mechanisms, GNNs progressively refine molecular representations, starting from the initial molecular embeddings to produce graph-level representations of the entire molecule. In this figure, nodes and edges of different colors correspond to different atoms and chemical bonds. The blue arrow in the second graph illustrates the message passing between neighboring nodes and edges.}
    \label{fig:GNN}
\end{figure}

Graph neural networks (GNNs) are specialized neural networks designed to handle graph-structured data~\cite{wu2020comprehensive}.
The core principle of GNNs lies in ``message passing'', where the nodes iteratively update their representations by exchanging information with neighboring nodes~\cite{scarselli2008graph}.
Given that molecular structures naturally map to graph structures, GNNs are a fitting choice for molecular modeling.

In this approach, atoms and bonds are mapped to nodes and edges, respectively (Fig.~\ref{fig:GNN}). Each node (atom) is associated with a feature vector encoding properties such as atom type, charge, etc., while edges (bonds) can also carry features, such as bond type or length.  
During training, a GNN learns to represent graph-structured data by applying multiple layers of message passing, where each node updates its representation based on information from its neighboring nodes and connecting edges.

After several iterations of message passing, the model produces rich representations for each node. 
For graph-level tasks, such as molecular property prediction, these node embeddings are aggregated to form a comprehensive graph-level representation of the molecule. This enables GNNs to effectively capture both local and global structural information, making them powerful tools for molecular modeling.

Building on the commonly used GNN architectures, such as {\em Graph Convolutional Networks (GCN)} \cite{kipf2017semi}, {\em Graph Attention Networks (GAT)}~\cite{velivckovic2017graph} and \\
{\em Graph Sample and Aggregation (GraphSAGE)}~\cite{hamilton2017inductive}, various models have been developed to represent molecular structures~\cite{kojima2020kgcn,liu2022attention,LIU2023106524}. Their effectiveness has been extensively validated using benchmark\\
datasets such as QM9 and BBBP~\cite{zhou2023unimol}, further demonstrating their utility in advancing AI-driven chemistry research.

The methods introduced above are designed for 2D molecular structures, whereas ``3D-GNNs'' more efficiently leverage molecular 3D information by incorporating spatial relationships derived from the atomic coordinates~\cite{reiser2022graph, godwin2022simple}. 
Instead of solely relying on the connectivity defined by chemical bonds, these models integrate geometric features such as interatomic distances, angles, and torsions. By embedding this spatial information, 3D-GNNs capture finer details about molecular interactions and conformations, making them particularly suitable for tasks involving quantum chemical properties, protein-ligand interactions, and materials discovery.

GNNs can also leverage large amounts of unlabeled data to learn representations with better generalization capabilities.
\textbf{Contrastive Learning}, a type of self-supervised learning technique, focuses on learning meaningful representations by comparing similar and dissimilar pairs of data points~\cite{chen2020simple}.
The central idea is to encourage the model to bring similar instances (called ``positive'' pairs) closer in the representation space while pushing dissimilar instances (called ``negative'' pairs) further apart.
The construction of positive and negative molecular pairs allows GNNs to more accurately capture the underlying patterns in molecular space. 
Feng et al.~\cite{feng2024unicorn} introduced ``Unicorn'', a unified pre-training framework for molecular representation. 
Unicorn addresses the limitations of existing pre-training approaches, which are often tailored to specific downstream tasks, by providing a more versatile solution.
It employs contrastive learning to align molecular views at three levels: 2D graph masking, 3D graph denoising, and 2D-3D mapping, offering a comprehensive understanding of molecule structures. 
This unified approach achieves state-of-the-art performance across a wide range of quantum, physicochemical, and biological tasks, showcasing its effectiveness as a general-purpose model for molecular representation.

\subsection{Transformer Based Representation Learning}

\begin{figure}[t]
    \centering
    \includegraphics[width=0.8\textwidth]{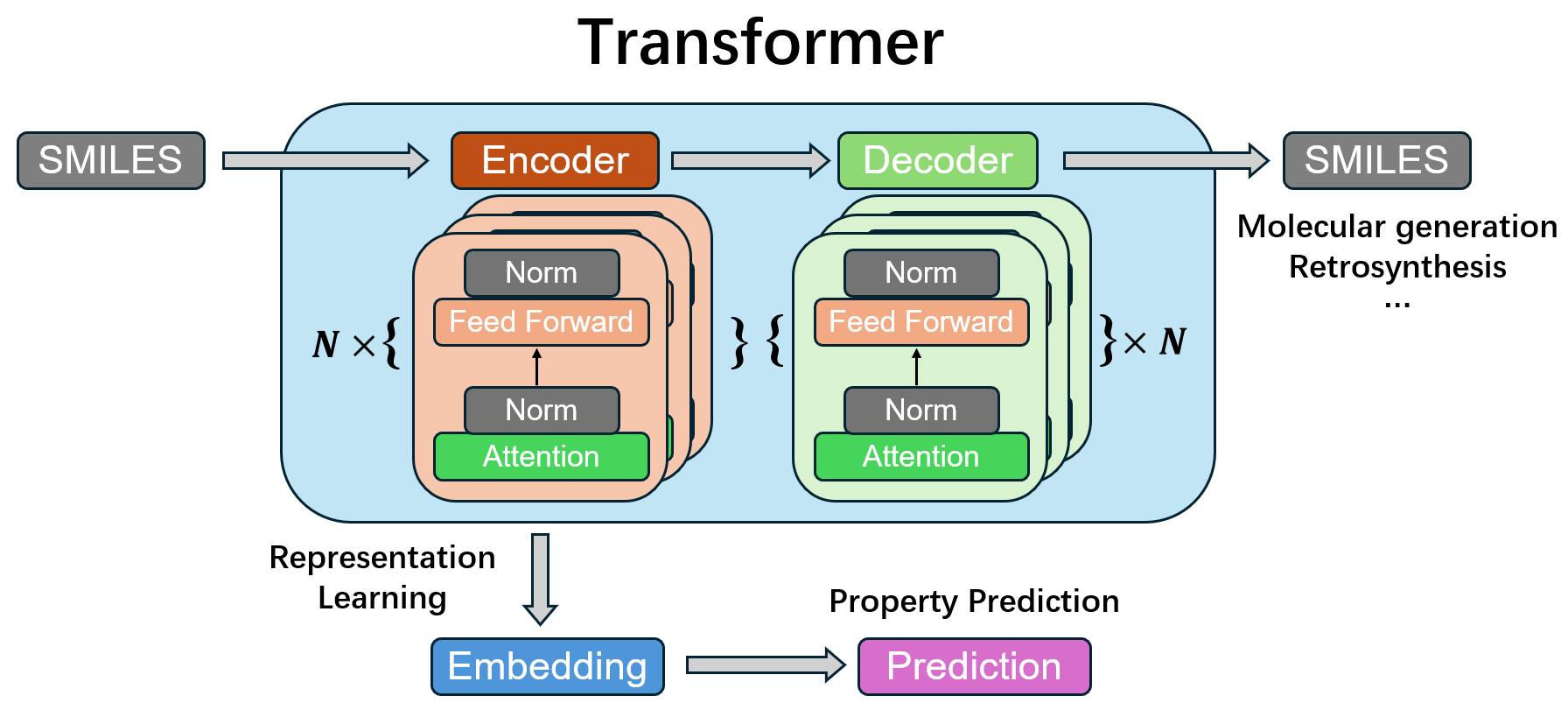}
    \caption{\textbf{Simplified transformer model in chemistry.} 
    Based on SMILES, reactions and molecules can be represented in text form. The Transformer architecture can perform various chemical tasks such as representation learning, property prediction, and retrosynthesis analysis by processing SMILES.}
    \label{fig:Transformer}
\end{figure}

Originally developed for natural language processing, Transformers have been adapted for chemical representation using sequential input of molecules~\cite{mao2021molecular}. 
Transformers excel at handling sequential data, making them particularly effective for processing SMILES, which are linear encodings of molecular structures. 
These models can capture long-range dependencies within sequences, which is important for understanding complex chemical properties.
Typically, Transformer models are pre-trained on large datasets and then fine-tuned for specific tasks such as property prediction or molecule generation~\cite{wang2019smiles, schwaller2019molecular} (Fig.~~\ref{fig:Transformer}).

Generative models based on the Transformer architecture can perform SMILES-to-SMILES tasks and, after fine-tuning for specific downstream applications, excel in tasks such as retrosynthesis and molecular design~\cite{tetko2020state, mazuz2023molecule}.  
Leveraging frameworks like {\em Bidirectional Encoder Representations from Transformers (BERT)} ~\cite{devlin2018bert}, Transformer can also handle various practical classification and prediction tasks in chemistry, such as reaction yield prediction~\cite{schwaller2021prediction, schwaller2019molecular}.  
Schwaller et al.~\cite{schwaller2021mapping} proposed the {\em Reaction Fingerprint (RXNFP)}, a model that generates fingerprints for reactions via BERT.
Their model employs attention mechanisms to focus on key atoms and reagents relevant to each reaction class.
Moreover, these reaction fingerprints are independent of the number of molecules involved in the reactions, making them versatile and applicable across diverse reaction datasets.
In this article, the authors further utilized RXNFP to embed and visualize reaction representations from the public USPTO-50k dataset~\cite{schneider_development_2015}, which comprises 50000 reactions categorized into 10 reaction classes. 

\subsection{Additional Remarks on Chemical Representation}

Chemical representation plays a vital role in the application of machine learning to chemistry, as it directly imapcts a model's ability to understand the chemical world. Traditional chemical representation methods often rely on chemists' prior knowledge, which can limit the diversity and richness of the representation. 
In contrast, learning-based representation methods aim to uncover underlying patterns from large volumes of unlabeled molecular data, leading to improved performance across various tasks. 

However, representations derived through training are often not explicitly linked to chemical structures or properties, leading to challenges in interpretability.
As AI continues to advance in the field of chemistry, we envision future representations combining the strengths of both traditional and learning-based approaches. The development of interpretable representation learning methods holds significant potential to accelerate progress in chemical research and enhance our understanding of complex molecular systems.

\section{ML Models for Different Applications}

The primary purpose of chemical research is to explore the structure, properties, and transformations of matter. However, traditional experimental and analytical methods in chemistry are often costly, posing significant challenges to the development of new materials and drugs. Here, we introduce several specific AI-driven applications in chemistry, including various prediction tasks, molecular design, and retrosynthesis. The integration of AI technologies has proven to significantly enhance the efficiency of these traditionally labor-intensive and time-consuming processes. 

\subsection{Prediction Problems}

\begin{figure}[t]
    \centering
    \includegraphics[width=\linewidth]{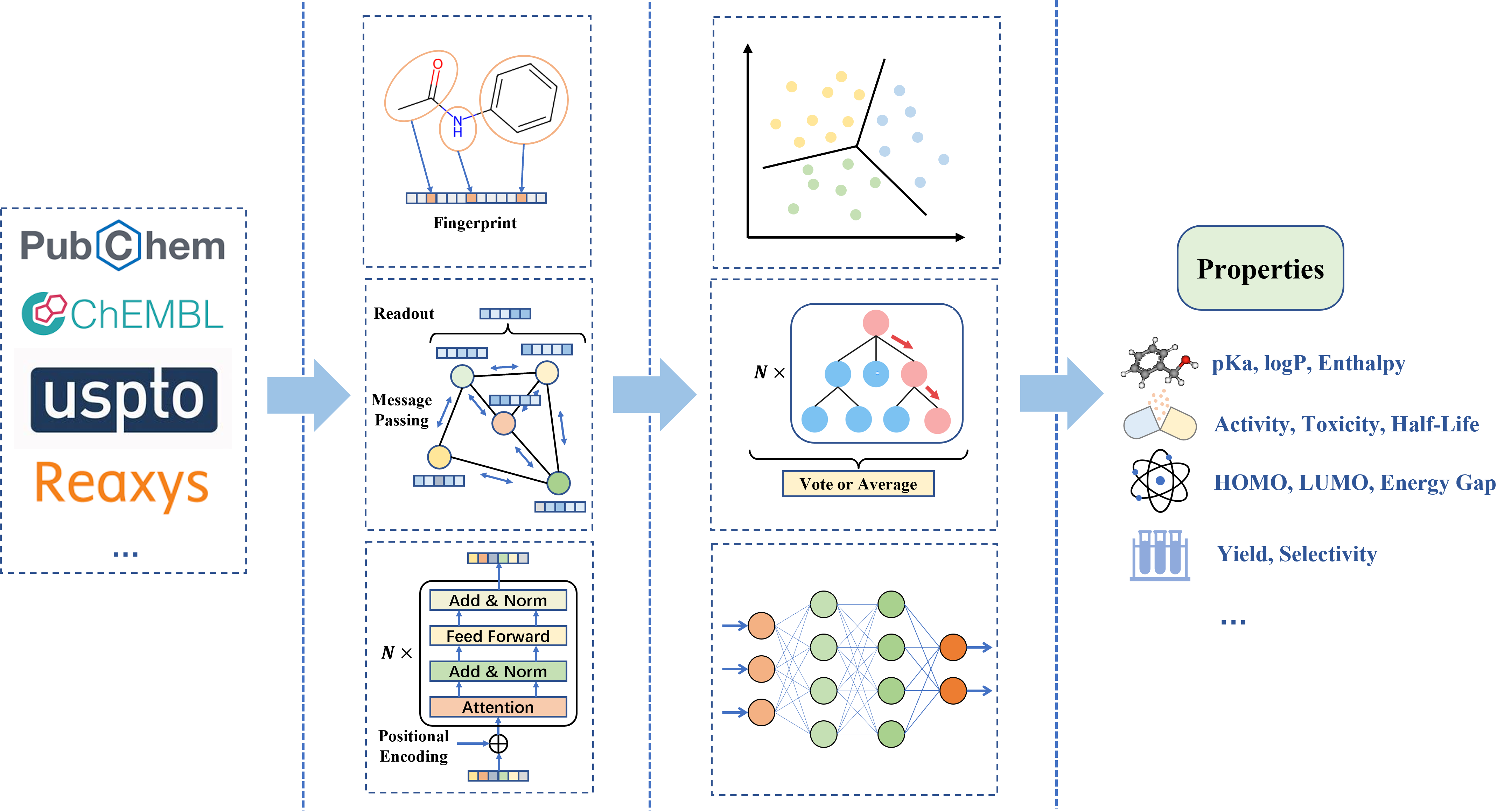}
    \caption{\textbf{Overview of the prediction problems in  chemistry aided by AI.} The overall framework begins with data, encompassing various molecular and reaction datasets (first column). Subsequently, different representation methods can be employed, such as molecular fingerprints, GNN-based, or Transformer-based representations (second column). Building on these representations, various machine learning models (third column) are utilized to predict a wide range of chemical properties (fourth column).}
    \label{fig:prediction}
\end{figure}

Many tasks in chemistry involve the prediction of certain features (Fig.~\ref{fig:prediction}). For example, in drug screening, chemists evaluate properties such as the {\em activity, toxicity,} and {\em half-life} of candidate compounds. In chemical synthesis, chemists prioritize reactions with high \emph{yield} and \emph{selectivity}. From a ML standpoint, these tasks can be framed as problems of ``\textbf{classification}'' (discrete output) or ``\textbf{regression}'' (continuous output), both of which are closely related to the representation of input data. Traditional ``non-deep learning'' methods, such as {\em random forests} or {\em support vector machines}~\cite{ahneman2018predicting}, typically feed molecular fingerprints or descriptors into their machine learning models. In contrast, recent advancements in the field increasingly leverage neural networks to learn molecular or reaction representations directly. These learned representations are then used to make predictions, either through additional network layers or traditional machine learning models~\cite{probst2022reaction}.

There are two primary approaches for modeling inputs in this realm. \textbf{The first approach} models each molecule as a graph,  and represents each reaction as a set of graphs, encompassing reactants, products, and other related components; then it learns the mappings from inputs to outputs via specific graph-based methods. Coley et al.~\cite{coley2019graph} adopted the {\em Graph Convolutional Neural Network (GCN)} to predict potential molecular sites of reactivity by calculating likelihood scores for each bond change between each atom pair. Candidate products are generated by enumerating the most likely changes, and then they are ranked by using another GCN to identify the final predicted product species. Zang et al.~\cite{zang2023hierarchical} proposed a hierarchical GNN framework for property prediction, capturing multi-scale information at the levels of atoms, motifs and entire molecules. Kwon et al.~\cite{kwon2022uncertainty} combined GNN with uncertainty-aware learning, enabling the simulatneous prediction of the mean and variance of reaction yields. Saebi et al.~\cite{saebi2023use} developed a hybrid framework that combines GNN with descriptor-based models; the predictions from both sources are integrated using an additional linear layer, and then the framework can produce the final predicted yield.

\textbf{The second approach} leverages the SMILES of molecules as inputs and employs sequence-based methods. Wang et al.~\cite{wang2019smiles} proposed a two-stage transform-based model called {\em SMILES-BERT}. This model was pre-trained on a masked SMILES recovery task using large-scale unlabeled data,  and later fine-tuned for different property prediction tasks, such as predicting  molecular activity. Similar strategies have also been applied to reaction-level predictions, such as yield prediction, where the input consists of the concatenated SMILES strings of reaction components. Yield-BERT, developed by Schwaller et al.~\cite{schwaller2021prediction}, was a pioneering work that utilized BERT~\cite{devlin2018bert} to encode reactions and make predictions. Later, their subsequent studies revealed that the performance could be further enhanced through data augmentations, achieved by permuting the order of reaction components~\cite{schwaller2020data}. Yin et al.~\cite{yin2024enhancing} extended this line of research by introducing an additional contrastive learning module into the BERT framework, leading to a more robust understanding of the input SMILES strings.

For prediction tasks, uncovering the specific relationships between molecular structures and chemical properties provides valuable insights into the underlying principles of chemistry. For example, Zheng et al.~\cite{zheng2019identifying} adopted the attention mechanism to identify these relationships. By analyzing the weights in the attention matrix, their proposed method can detect which segments in SMILES strings are prioritized and which are disregarded. This enables the identification of molecular fragments that influence specific properties. In their experiments for toxicity prediction, this approach successfully identified the functional groups, such as phosphoric acid esters, aliphatic halides, that contribute to toxicity. Recently, Wong et al.~\cite{wong2024discovery} developed a \emph{Monte Carlo Tree Search (MCTS)} based approach to identify substructures associated  with high antibiotic activity. They first trained an ensemble of predictive models, provided by Chemprop package~\cite{heid2023chemprop}, to predict the activity of a given substructure. The substructure was then iteratively pruned using MCTS, with deletions selected based on high prediction scores. The identified substructures offer promising avenues for exploring novel structural classes, and thus the method has the potential to play a pivotal role in developing antibiotics to address the growing resistance crisis. 

\begin{table}[!h]
    \centering
    \caption{Summary of recent studies on prediction problems.}
    \begin{tabular}{ccp{2.5cm}cp{8.5cm}}
        \toprule
        Level & Targets & \multicolumn{1}{c}{Methods} & Graph/Sequence & \multicolumn{1}{c}{Highlights} \\
        \midrule
        Molecule & Property & \cite{zheng2019identifying} & Sequence & Applying a self-attention mechanism with BiLSTM to enhance model interpretability. \\
        & & \cite{wang2019smiles} & Sequence & Pretraining the model through a masked SMILES recovery task. \\
        & & \cite{lu2019molecular} & Graph & Incorporating hierarchical layers to capture features from multilevel interactions.\\
        & & \cite{zang2023hierarchical} & Graph & Encoding hierarchical information from node-motif-graph structures and introducing five tasks for self-supervised pre-training. \\
        & &  \\
        Reaction & Products & \cite{coley2019graph} & Graph & Dividing the task into two stages of reactivity perception and outcome scoring. \\
        & & \cite{chen2022generalized} & Graph & Deriving generalized reaction templates to only describle the net changes in electron configurations and proposed a network to predict products based on them. \\
        & & \cite{lu2022unified} & Sequence & Developing a unified model to tackle multiple reaction prediction tasks with task-specific prompts. \\
        & Yield & \cite{schwaller2021prediction} & Sequence & Utilizing BERT to encode reactions and make prediction. \\
        & & \cite{schwaller2020data} & Sequence & Applying molecule permutations and SMILES randomizations to augument the data. \\
        & & \cite{kwon2022uncertainty} & Graph & Adapting a GNN and uncertainty-aware learning and inference to predict yield and variance together. \\
        & & \cite{saebi2023use} & Graph & Integrating structure-based and chemical-feature-based predictions. \\ 
        & & \cite{yin2024enhancing} & Sequence & Utilizing masked language modeling and reaction-condition-based contrastive learning to pretrain the BERT. \\
        \bottomrule
    \end{tabular}
    
    \label{tab:prediction}
\end{table}

\subsection{Molecular Design}

\begin{figure*}[!h]
    \centering
    \includegraphics[width=0.8\linewidth]{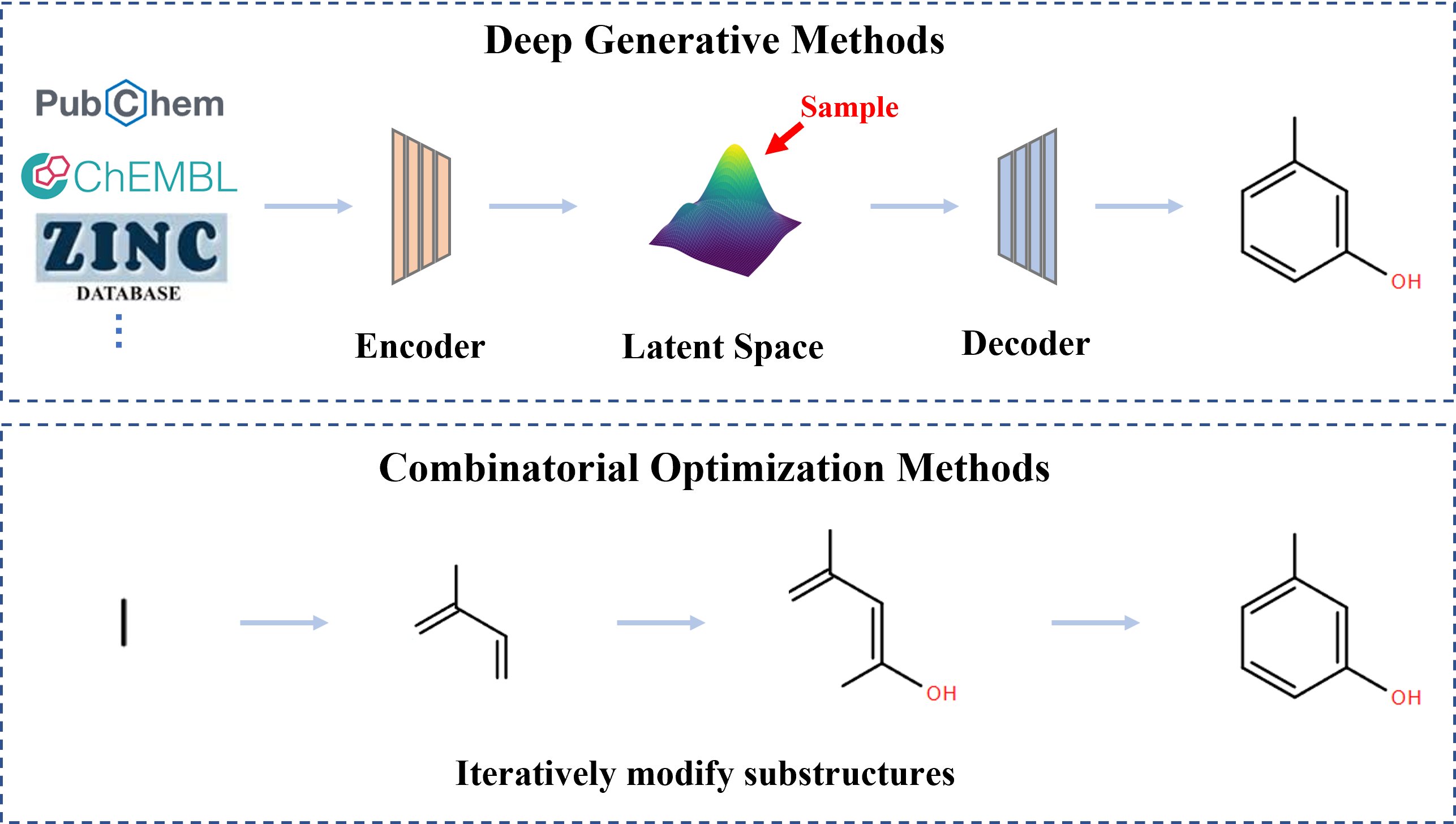}
    \caption{\textbf{Overview of the molecular design methods.}}
    \label{fig:mol_design}
\end{figure*}

The discovery and design of molecules with desired properties is one of the fundamental tasks of chemical research. For instance, researchers aim to develop molecules that can selectively bind to specific biological targets or design catalysts with tailored structures. Early studies predominantly focused on {\em virtual screening}, where the properties are predicted for molecules within large chemical libraries, followed by experimental validation of the most promising candidates~\cite{shoichet2004virtual}. For example, Stokes et al.~\cite{stokes2020deep} performed virtual screening on more than 107 million molecules from several chemical libraries, which led to the discovery of the first AI-discovered antibiotic, ``halicin''. However, the effectiveness of virtual screening is inherently constrained by the coverage of chemical libraries. Expanding library size may improve coverage but significantly increases time costs for property prediction. To address these limitations, \textbf{\textit{de novo} molecular design} has emerged as a compelling alternative. Unlike virtual screening, this approach does not rely on  existing molecular libraries but instead generates novel molecules tailored to specific properties. Current methods for \textit{de novo} molecular design can be broadly categorized into two main approaches: deep generative methods and combinatorial optimization methods (Fig.~\ref{fig:mol_design}). 

\subsubsection{Deep Generative Methods}

Recent advances in deep generative models have yielded many exciting results~\cite{achiam2023gpt, brooks2024video}, making them an increasingly popular approach for \textit{de novo} molecular design. Drawing from the advancements in neural language processing, a widely explored strategy is to generate SMILES sequences for molecules. For example, Segler et al.~\cite{segler2018generating} adopted recurrent neural network (RNN) to generate a large number of unconstrained molecular SMILES, which can serve as candidates for virtual screening.

From a practical perspective, it is often essential for generated molecules to exhibit specific properties, prompting the development of various efforts towards conditional generation. Olivecrona et al.~\cite{olivecrona2017molecular} introduced a reinforcement learning method to tune a pre-trained RNN for generating desirable compounds. Kang et al.~\cite{kang2018conditional} trained a Variational AutoEncoder (VAE) that integrates property prediction with molecular generation. The VAE is a generative model consists of an encoder that maps data into a latent space and a decoder that reconstructs data from embeddings in the latent space~\cite{Kingma2019vae}. In their study, they generated new molecules by sampling from the conditional generative distribution of the decoder. Mao et al.~\cite{mao2023transformer} explored the relationship between the IUPAC name---a standardized chemical naming system proposed by IUPAC---and molecular structure via a transformer-based network, enabling the generation of desired molecular structures through IUPAC name editing. G\'{o}mez-Bombarelli et al.~\cite{gomez2018automatic} performed gradient-based optimization in the continuous embedding space to search for optimized representations, which were then converted into molecules using a decoder. However, the decoding process can sometimes produce invalid molecules, especially when the targets lie far from the training data in the embedding space. To mitigate this limitation, Griffiths et al.~\cite{griffiths2020constrained} introduced a validity constraint to improve the reliability of the generated molecules. 

Unlike SMILES-based generation, Liu et al.~\cite{liu2018constrained} presented a graph-based VAE to generate moleuclar graph structures. They employed a GNN as the encoder, while the decoder follows a generative procedure that iteratively selects edges to construct the molecular graph. Jin et al.~\cite{jin2018junction} also proposed a method for directly generating molecular graphs. They trained a GNN that combines molecular graphs with extracted tree scaffolds, first predicting the scaffold and then recovering the graphical structure. This recovery procedure is akin to building piece-by-piece from a set of valid components rather than atom-by-atom, which may involve chemically invalid intermediates. Subsequently, Jin et al.~\cite{jin2020hierarchical} proposed another hierarchical encoder-decoder to learn the patterns of atoms, attachments and motifs. The decoder generates the target molecule in a coarse-to-fine manner: selecting a motif, predicting the attachment site, and determining the points of contact with the current molecule. 

Recently, diffusion models have gained widespread attention in the field of image generation. In brief, diffusion models are a class of generative models that iteratively transform random noise into structured data, learning the underlying data distribution by reversing a diffusion process~\cite{yang2023diffusion}. Researchers have increasingly begun experimenting with these models for molecular generation, particularly focusing on the 3D structures of molecules. For example, Hoogeboom et al.~\cite{Hoogeboom2022Equivariant} introduced a diffusion model capable of generating 3D molecular structures while maintaining E(3)-equivariant property, meaning it is isotropic to rotation, translation and reflection. Igashov et al.~\cite{igashov2024equivariant} also proposed an E(3)-equivariant diffusion model for designing molecular linkers between drug fragments. Their model first predicts the size of the possible linker and then iteratively updates atom types and coordinates through the denoising network conditioned on the input fragments.

\subsubsection{Combinatorial Optimization Methods}
Another approach is to formulate the molecular design as a discrete optimization task, where substructures of molecules are modified based on guidance by some oracle. One widely used method in this context is the {\em genetic algorithm (GA)}, inspired by the process of biological evolution~\cite{sheridan1995using, michalewicz1996evolutionary}, which has demonstrated promising performance in molecular design. In general, a GA maintains a population of ``candidate solutions'' and iteratively applies genetic operations (e.g., crossover and mutation) among the candidates to generate new solutions. The key elements of a GA include an appropriate definition of the genetic operations and an effective fitness function that guides the evolutionary process in the desired direction~\cite{katoch2021review}.

Jenson et al.~\cite{jensen2019graph} developed a graph-based GA that performs crossovers and mutations directly on molecular graphs. They designed a list of genetic operations, including ``append atom'', ``insert atom'', ``change atom type'' and so on, demonstrating  performance comparable with previous methods based on deep learning~\cite{yang2017chemts}. Subsequently, Nigam et al.~\cite{nigam2019augmenting} introduced a neural network enhanced GA for molecular design. By integrating a discriminator to adaptively model the penalty term in the fitness function, their approach facilitated the generation for more diverse molecules. Ahn et al.~\cite{ahn2020guiding} utilized a \emph{Long-Short Term Memory (LSTM)} network to learn mutation rules dynamically, replacing fixed genetic operators. In another study, Nigam et al.~\cite{nigam2022parallel} proposed a parallel GA framework that simultaneously propagates two populations---one focused on exploration and the other on exploitation. These populations exchange members in each iteration to achieve a balance between diversity and refinement. Fu et al.~\cite{fu2022reinforced} reformulated the evolutionary process as a Markov decision process, leveraging a neural network to make informed decisions, thereby replacing traditional random-walk-like exploration.

\begin{table}[!b]
    \centering
    \caption{Summary of recent studies on molecular design.}
    \begin{tabular}{p{3cm}ccp{10cm}}
        \toprule
        \multicolumn{1}{c}{Methods} & Methodology & Architecture & \multicolumn{1}{c}{Highlights} \\
        \midrule
        \cite{segler2018generating} & Gen. & RNN & The first sequence-to-sequence model for molecular generation. \\
        \cite{olivecrona2017molecular} & Gen.+RL & RNN & Introducing a reinforcement learning method to tune the pretrained RNN for generating structures with certain specified properties. \\
        \cite{kang2018conditional} & Gen. & VAE & Simultaneously performing both property prediction and molecule generation for conditional design. \\
        \cite{mao2023transformer} & Gen. & Transformer & Establishing the relationship between IUPAC name and molecular structure, enabling molecular design via editing the IUPAC name. \\
        \cite{gomez2018automatic} & Gen. & VAE & Embedding molecules into a continuous space, where optimization is performed to find molecules with desired property. \\
        \cite{griffiths2020constrained} & Gen. & VAE & Reformulating the search procedure as a constrained Bayesian optimization problem, enhancing the validity of generated molecules. \\
        \cite{liu2018constrained} & Gen. & VAE & Generating molecular graphs via a sequential graph extension manner. \\
        \cite{jin2018junction} & Gen. & VAE & Generating a tree-structured scaffold and assembling nodes to form a complete molecule. \\
        \cite{jin2020hierarchical} & Gen. & VAE & Proposing a hierarchical graph encoder-decoder for molecule generation. \\
        \cite{Hoogeboom2022Equivariant} & Gen. & Diffusion & Introducing an equivariant denoising diffusion model that jointly operates on atom coordinates and atom type. \\
        \cite{igashov2024equivariant} & Gen. & Diffusion & An E(3)-equivariant diffusion model for generating molecular linkers between drug fragments. \\
        \cite{jensen2019graph} & GA & - & Adopting a genetic algorithm at the molecular graph level for molecular design. \\
        \cite{nigam2019augmenting} & GA & MLP & Applying a discriminator to adaptively model the penalty, enhancing the diversity of generated molecules. \\
        \cite{fu2022reinforced} & GA+RL & GNN & Using reinforcement learning to guide the selection of targets for genetic operation. \\
        \cite{you2018graph} & RL & GCN & Combining graph representation, reinforcement learning and adversarial training in a unified framework. \\
        \cite{zhou2019optimization} & RL & MLP & Techniques including double Q-learning and randomized value functions are employed for multi-objective molecular design. \\
        \cite{gottipati2020learning} & RL & MLP & Treating the generation of molecular structures as a sequential decision process of selecting reactants and reactions. \\
        \cite{fu2022differentiable} & Com. Opt. & GNN & Proposing differentiable scaffolding tree and enabling the gradient-based optimization on chemical graph structure. \\
        \bottomrule
    \end{tabular}

    \label{tab:molecule}
\end{table}

{\em Reinforcement learning (RL)} is a computational paradigm that focuses on enabling agents to make sequential decisions within dynamic environments, with the goal of maximizing cumulative rewards \cite{arulkumaran2017deep}. In the context of molecular design, the generation of molecules can be regarded as a series of graph expansion operations, where RL is used to learn the decision-making process. In this setup, the action space typically consists of operations such as adding atoms or substructures, while the state space encompasses the set of all molecules that can be constructed through these actions. For example, You et al.~\cite{you2018graph} proposed a unified framework integrating molecular graph representation, reinforcement learning and adversarial training to facilitate molecule generation. Zhou et al.~\cite{zhou2019optimization} developed a value function based multi-objective RL model, which is capable of starting from scratch without relying on pre-training. In contrast to the aforementioned setups, Gottipati et al.~\cite{gottipati2020learning} modeled molecule generation as a series of synthetic sequences. In this formulation, selecting reactants constitutes an action, while the resulting product molecule represents a state of the system. Although this setup introduces additional complexity, it ensures the synthesizability of generated molecules, as the decision sequences directly correspond to feasible synthetic pathways.

In addition to the aforementioned approaches, Fu et al.~\cite{fu2022differentiable} proposed a differentiable formulation for the molecular scaffolding tree and employed a GNN to approximate the property oracle.  
Namely, they transformed the original discrete local search into a differentiable optimization problem, so as to enable the use of gradient-based algorithms to directly optimize molecular structures, significantly enhancing the efficiency of the design process.

\subsection{Retrosynthesis}

\begin{figure}[!b]
    \centering
    \includegraphics[width=\linewidth]{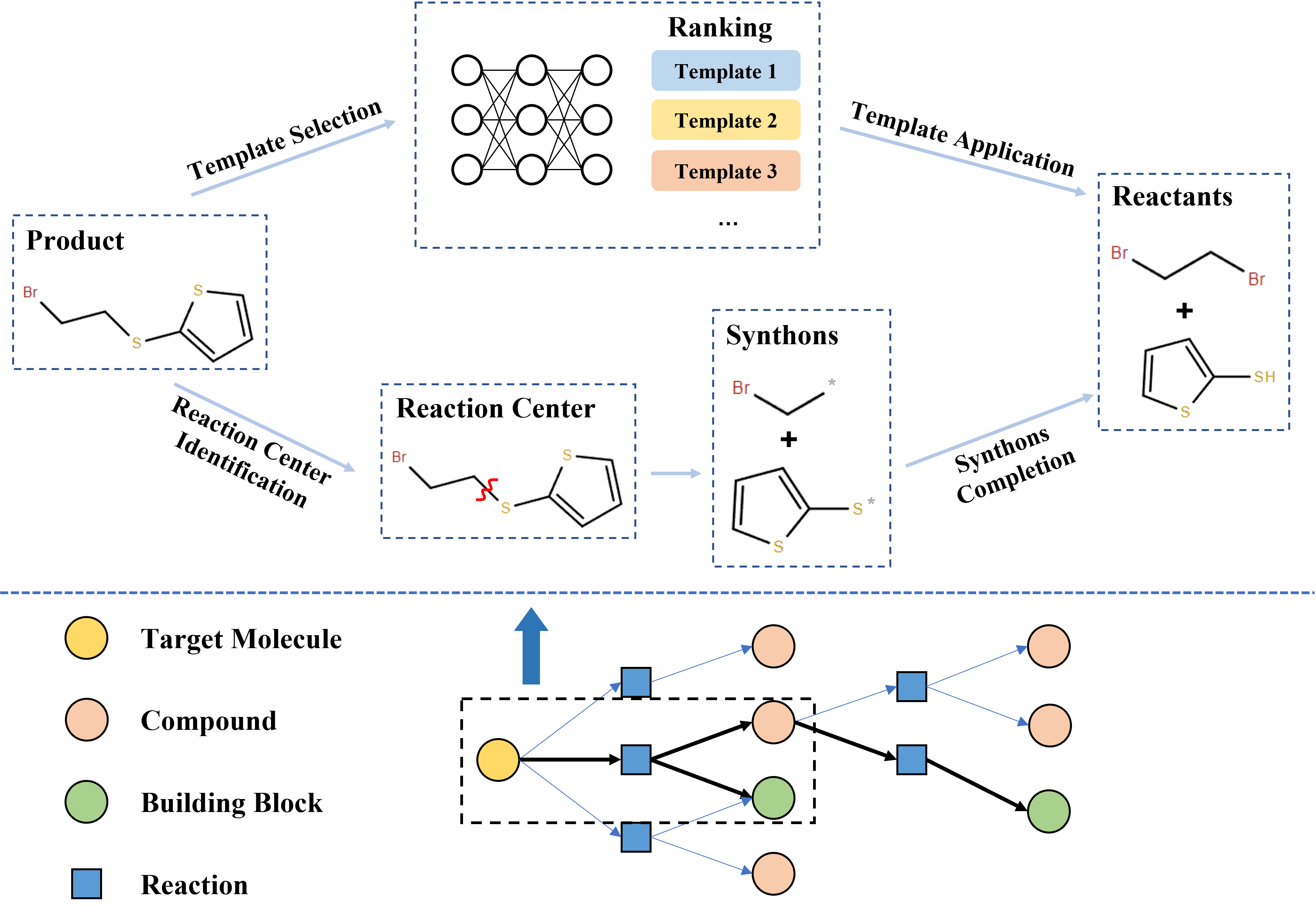}
    \caption{\textbf{Overview of the retrosynthesis task.} In single-step retrosynthesis, the task is to identify the reactants corresponding to a target product. In multi-step retrosynthesis, the goal is to construct a complete synthesis tree where all leaf nodes correspond to available chemicals (building blocks). Common retrosynthesis methods are typically categorized into template-based and template-free approaches.}
    \label{fig:retro}
\end{figure}

Retrosynthesis focuses on designing synthetic routes for target molecules, serving as a critical step in bridging molecular designs and practical applications. For instance, newly designed drugs or discovered materials require carefully devised synthetic pathways to achieve industrial-scale production. Existing methods for retrosynthesis can be broadly categorized into two types: ``template-based'' and ``template-free'' (Fig.~\ref{fig:retro}), where \textbf{reaction templates} are the predefined rules that describe the transformation patterns from reactants to products. Template-based methods identify the most suitable template for the target molecule from a set of given templates~\cite{coley2017computer}. On the other hand, template-free methods leverage generative models to directly predict reactants, bypassing the need for predefined templates~\cite{liu2017retrosynthetic}.

\subsubsection{Template-based methods}

``Synthia'' (formly known as ``Chematica'') is a hybrid expert-AI system which incorporates comprehensive reaction rules with advanced computational techniques, including random forests, heuristic functions and state-of-the-art neural networks, to facilitate synthesis planning~\cite{mikulak2020computational}. One of the earliest attempts to apply deep learning in retrosynthesis was by Segler et al.~\cite{segler2017neural}, who utilized neural networks to learn patterns of the molecular functional group and predict the most probable transformation rules for target products. Building on this, they incorporated their approach with MCTS to successfully extend it to multi-step synthesis planning~\cite{segler2018planning}. Dai et al.~\cite{dai2019retrosynthesis} proposed modeling retrosynthesis as a probability reasoning problem and then introduced a Conditional Graph Logic Network to solve it. Additionally, Chen et al.~\cite{chen2021deep} observed that most elementary reaction changes occur locally and then developed a GNN-based model to learn localized reaction template effectively. 

Although template-based approaches offer results with good interpretability, they are subject to some limitations. The effectiveness of these methods heavily depends on the quality of the given template set. Heid et al.~\cite{heid2021influence} investigated how the factors such as template generality, canonicalization, and exclusivity influence the performance of different template-based models. Their findings revealed that duplicate and nonexclusive templates could significantly reduce model accuracy. Therefore, they developed a canonicalization and hierarchical template correction algorithm to mitigate this limitation. It is also argued that template-based models do not truly learn the underlying chemical mechanisms but instead encode synthesis rules~\cite{schwaller2020predicting}. Furthermore, the reliance on a limited set of training templates constrains their ability to discover novel reactions~\cite{segler2017modelling, yan2020retroxpert}. To overcome this issue, Yan et al.~\cite{yan2022retrocomposer} attempted to compose templates from a predefined set of template building blocks, which has the potential to discover new reaction templates from the training templates set.

\subsubsection{Template-free methods}

Distinct from template-based methods, template-free approaches directly learn patterns from target molecules and reactants, offering greater potential for generalizability. An early template-free model for retrosynthesis was introduced by Liu et al.~\cite{liu2017retrosynthetic}, who successfully employed a seq2seq model to directly predict the SMILES strings of reactants for a given product. Later, Zheng et al.~\cite{zheng2019predicting} implemented a retrosynthetic reaction predictor based on the Transformer architecture. By coupling with an additional molecular syntax corrector, they could fix syntax errors in raw candidate reactants, leading to more accurate and reasonable predictions. Chen et al.~\cite{chen2019learning} constructed numerous auxiliary targets for pre-train by random bond breaking or template-based transformations. They further incorporated a latent variable model into the framework to capture reaction types thus producing diverse predictions. In contrast to the previous methods, which primarily relied on sequential information, Mao et al.~\cite{mao2021molecular} proposed a ``Graph Enhanced Transformer'' framework that incorporates molecular graphical information with sequential representations. This hybrid approach significantly improves the ability to capture structural and sequential patterns, broadening the scope of template-free retrosynthesis. 

Although the direct prediction methods mentioned above are both simple and effective, they deviate from the \textbf{deductive reasoning process} typically employed by chemists. In particular, chemists often approach retrosynthesis by iteratively fragmenting the target molecule into idealized molecular fragments, known as ``synthons'', based on reactivity. The sites in the target molecule with high reactivity are referred to as ``reaction center'', which are typically the bonds that will break during the reaction. Therefore, some studies try to reformulate retrosynthesis as a two-stages process: first, splitting the target molecule into intermediate synthons by identifying reaction centers; second, generating the reactants associated with these synthons.
Shi et al.~\cite{shi2020graph} first implemented this strategy via a graph-to-graph model. Meanwhile, Yan et al.~\cite{yan2020retroxpert} proposed an alternative architecture and introduced a novel data augmentation technique that incorporated unsucessfully predicted synthons, thereby enhancing the model's robustness. Somnath et al.~\cite{somnath2021learning} simplified the synthon completion step by precomputing a vocabulary called ``leaving groups'', which records the subgraphs differing between synthons and their corresponding reactants. This transformation recasts the synthon completion step from a generation problem into a classification problem, significantly reducing its complexity. More recently, Han et al.~\cite{han2024retrosynthesis} redefined the retrosynthesis problem as a string editing task. Their approach begins by identifing the reaction center through repositioning within the SMILES of the target molecule. Potential attachment sites are then marked with placeholders, followed by token decoding to complete the synthons. This new approach bridges the gap between string-based and graph-based methods, offering an efficient alternative for retrosynthesis.

\begin{table}[!h]
    \centering
    \caption{Summary of recent studies on retrosynthesis.}
    \begin{tabular}{cp{3cm}p{10cm}}
        \toprule
        Type & \multicolumn{1}{c}{Methods} & \multicolumn{1}{c}{Highlights} \\
        \midrule
        Template-Based & \cite{mikulak2020computational} & A hybrid expert-AI system incorporated reaction templates with several ML models. \\
        & \cite{segler2017neural} & Introducing deep neural networks to prioritize the most suitable templates. \\
        & \cite{dai2019retrosynthesis} & Formulating the task as a probability reasoning problem. \\
        & \cite{chen2021deep} & Proposing a local retrosynthesis framework to learn localized templates. \\
        & \cite{yan2022retrocomposer} & Composing templates from basic building blocks to generate new templates. \\
        & \cite{gao2022amortized} & Generating synthetic routes as trees conditioned on a target molecular embedding. \\
        & \cite{luo2024projecting} & Introducing a sequential representation for synthetic routes and learning via  Transformer. \\
        
        Template-Free & \cite{liu2017retrosynthetic} & Treating the retrosynthesis task as a sequence-to-sequence mapping problem. \\
        & \cite{zheng2019predicting} & Introducing an additional molecular syntax corrector for post-processing. \\
        & \cite{chen2019learning} & Augmenting training data by random bond breaking or template transformations. \\
        & \cite{mao2021molecular} & Incorporating molecular graphical information with sequential representations. \\
        & \cite{shi2020graph} & Firstly implementing the two-stages strategy via a graph-to-graph model. \\
        & \cite{yan2020retroxpert} & Incorporating unsuccessfully predicted synthons to augment data. \\
        & \cite{somnath2021learning} & Precomputing a leaving groups to simplify the synthon completion step. \\
        & \cite{han2024retrosynthesis} & Redefining the retrosynthesis as a string editing task. \\
        \bottomrule
    \end{tabular}

    \label{tab:retro}
\end{table}

\subsubsection{Multi-step retrosynthesis}
                    
In practical synthetic planning, the limited diversity of commercially available chemicals should be taken into account. Consequently, synthetic routes are often designed as multi-step processes, progressing from basic ``building blocks''---readily available chemicals used as starting materials---to intermediates, and ultimately culminating in the target molecules. 

A number of studies try to combine single-step retrosynthesis methods with various search algorithms to tackle the challenges of multi-step retrosynthesis. For example, Lin et al.~\cite{lin2020automatic} trained a Transformer model for an end-to-end retrosynthesis task and integrated it with MCTS to infer multi-step synthetic pathways. The search process starts from the target molecule and iteratively proceeds through four steps: selection, expansion, simulation, and backpropagation. Their single-step prediction model is utilized for node generation during the expansion step and for weighted molecule sampling during the simulation step. In addition to MCTS, other search strategies, such as beam search~\cite{schwaller2020predicting}, depth-first search~\cite{shibukawa2020compret}, A* search~\cite{chen2020retro, han2022gnn}, have also been widely explored.

Being different from the methods described above, which infer synthetic pathways in a reverse manner starting from the target molecule, there  also exist several approaches that directly construct pathways in a forward manner, beginning from basic building blocks and progressing toward the target molecule. Gao et al.~\cite{gao2022amortized} modeled synthetic pathways as synthetic trees and formulated a Markov decision process to the construction of these trees. In their model, the root molecule(s) of an intermediate synthetic tree are treated as the current state, and the decisions are made from possible actions at each iteration to iteratively build the synthetic tree via a bottom-up manner. More recently, Luo et al.~\cite{luo2024projecting} adopted the post-order traversal sequence to represent the synthetic tree. This sequential format simplifies the complexity of the tree structure and allows for learning via the prediction of the next token using a Transformer, similar to the tasks in natural language processing.

It is worth noting that the proposed methods from Gao~\cite{gao2022amortized} and Luo~\cite{luo2024projecting} are even capable of finding synthesizable analogs for some synthetically ``infeasible'' molecules, such as those generated by certain molecular design models. Since their methods are essentially based on template-driven reasoning, with each step adhering to a specific reaction template, they ensure the feasibility of the synthetic route. However, as every coin has two sides, this also limits their ability to discover novel reactions.

\subsection{{Integrating ML into Chemical Robots}}

{\em Self-Driving Laboratories (SDLs)} represent a groundbreaking innovation in scientific research, integrating automation, robotics, and machine learning to accelerate the discovery and optimization of new materials, molecules, and processes~\cite{tom_self-driving_2024}.
These systems are capable of autonomously conducting experiments, collecting data, analyzing results, and iterating on experimental designs---all without direct human intervention.
A key implementation of SDLs is the use of {\em AI Robots}, which are often
designed to mimic human work patterns, enabling efficient and scalable experimentation. 
This section introduces several representative AI Robots employed in SDLs.

In 2020, Cooper's group introduced a mobile  system  that is designed to autonomously conduct chemical experiments, with a focus on improving photocatalysts for hydrogen production from water~\cite{burger_mobile_2020}. 
Powered by a Bayesian search algorithm~\cite{rasmussen_gaussian_2005}, the robot conducted 688 experiments over the span of eight days, successfully identifying photocatalyst mixtures with six times the activity of the initial formulations.
Utilizing laser scanning and tactile feedback for precise positioning, the robot efficiently carried out tasks such as sample manipulation and instrument operation with high accuracy.
This approach 
offers a scalable and versatile solution for automating chemical research, with potential applications across a broad spectrum of material discovery challenges.

In addition, Luo and Jiang's group developed ``AI-Chemist''~\cite{10.1093/nsr/nwac190},
a system comprising three integrated modules: a machine-reading module that automatically analyzes extensive chemical literature to capture existing knowledge, a mobile robot module that performs various chemical experiments to generate experimental data, and a computational brain module that uses theoretical calculations to develop predictive models for key properties such as free energy change.
The combination of these three modules yields an all-round AI-Chemist with scientific reasoning capabilities.
The AI-Chemist has been tested in three distinct chemical tasks, demonstrating its competency.
The continued advancement of these all-encompassing AI-Chemists
has the potential to revolutionize the future of chemical laboratories.

The ``A-Lab'' is an autonomous platform designed to accelerate the discovery of novel inorganic materials by integrating robotics and machine learning techniques~\cite{szymanski_autonomous_2023}.
Over 17 days, it successfully synthesized 41 compounds from 58 targets.
The system utilized robotic stations for sample preparation, heating, and X-ray diffraction characterization, where the adopted ML models can further interpret the obtained diffraction patterns for analyzing the   outcomes.
It also employs active learning algorithms to refine synthesis recipes based on experimental results, directing the process toward high-yield targets.
This approach demonstrates the power of AI techniques  for automatic and 
high-throughput material discovery and optimization.

The AI-EDISON system~\cite{doi:10.1126/sciadv.abo2626} is an autonomous robotic platform designed for the exploration, discovery, and optimization of nanomaterials, specifically gold nanoparticles (AuNPs). 
This system integrates real-time spectroscopic feedback, machine learning algorithms, and a unique digital signature format to enable reproducible and efficient multistep synthesis.
By utilizing quality-diversity (QD) algorithms~\cite{10.1145/2001576.2001606, mouret_illuminating_2015}, which are optimization techniques designed to explore and generate a diverse set of high-quality solutions, AI-EDISON performs open-ended exploration across a high dimensional synthetic space, discovering diverse nanostructures.
The platform uses UV-Vis spectroscopy to
guide the synthesis process through extinction spectrum simulations, optimizing optical properties and achieve yield of up to 95\%.

\section{LLMs for Chemistry}

With the notable successes of large language models (LLMs) in natural language processing tasks~\cite{brown2020language, achiam2023gpt, touvron2023llama}, researchers have increasingly explored how to harness this powerful tool in the field of chemistry~\cite{m2024augmenting, boiko2023autonomous, jablonka2024leveraging, jablonka202314}. The relevant research of LLMs for chemistry can be broadly characterized by the following three categories.

The first category leverages
LLMs to create \textbf{chemical agents}. For example, ``ChemCrow''~\cite{m2024augmenting} integrates several expert-designed chemical tools with GPT-4 to streamline the reasoning process for common chemical tasks such as drug design and synthesis. Boiko et al.~\cite{boiko2023autonomous} also developed a multi-LLMs-based intelligent agent called ``Coscientist'', which is capable of executing complex chemical experiments. This agent can effectively search the Internet for related information on requested reactions and design protocols for mastering automation systems. Kang and Kim~\cite{kang2023chatmof} built an AI system focused on metal-organic framework, named ``ChatMOF''. In addition to an LLM agent and related toolkits, ChatMOF introduces another LLM as an evaluator to assess current outputs and make final responses.

The second category involves using LLMs to enhance downstream models or developing chemical domain-specific LLMs. Schwaller et al.~\cite{schwaller2021prediction} treated organic chemistry as a language, utilizing BERT to encode reaction SMILES as input for downstream yield prediction models. Bagal et al.~\cite{bagal2021molgpt} imitated the training method of LLMs and trained a mini GPT model for molecular generation by predicting the next SMILES token. ``ChemLLM''~\cite{zhang2024chemllm} is an open-source chemical LLM developed by the Shanghai AI Laboratory. This model is trained based on the InternLM2-Base-7B~\cite{cai2024internlm2}, and leverages 1.7M Multi-Corpus and 7M chemical data collected from online repositories for instruction tuning. The experimental results suggest that ChemLLM significantly outperforms general-purpose LLMs of similar scale and achieves results comparable to GPT-4 on core chemical tasks but with a much smaller model size.

The last category focuses on evaluating the chemical capabilities of LLMs. Guo et al.~\cite{guo2023can} established a comprehensive benchmark to assess the chemical knowledge of LLMs. The benchmark covers 8 practical chemistry tasks demanding different abilities, such as understanding, reasoning and explaining. Mirza et al.~\cite{mirza2024large} curated over 7000 question-answer pairs from university exams and chemical databases, covering most topics taught in undergraduate and graduate chemistry curricula. In conjunction with the development of ChemLLM, the Shanghai AI Laboratory constructed a chemical benchmark containing 4100 questions in the form of multiple-choice~\cite{zhang2024chemllm}. 

The evaluation results indicate that LLMs have outperformed conventional techniques in certain aspects of chemistry~\cite{jablonka2024leveraging, guo2023can}. Jablonka et al.~\cite{jablonka2024leveraging} investigated the performance of GPT-3 in predicting solid-solution formation in high-entropy alloys. With using only around 50 data points  for fine-tuning, they can achieve the performance comparable to the state-of-the-art methods specifically designed for this application (who used more than 1000 data points for training)~\cite{pei2020machine}. Mirza et al.~\cite{mirza2024large} developed a web application to survey chemists for their benchmark. A total of 41 chemists from different specializations participated in their tests, and the leading LLMs, Claude 3 and GPT-4, even surpassed the best scores of these experts. 

Despite of those remarkable performances, LLMs struggle with the tasks that require a precisely understanding of SMILES representation. Guo et al.~\cite{guo2023can} found that LLMs exhibit extremely low accuracy when converting between different molecular formats, such as SMILES to IUPAC name. Similar results were also reported in~\cite{castro2023large}, who discovered that LLMs often miss certain functional groups or introduce non-existent atoms in the representation. Forthermore, LLMs may generate molecules that are chemically unreasonable, which is a type of hallucination issues commonly observed in LLMs in the context of chemistry~\cite{guo2023can}. Eliminating hallucination in LLMs, particularly in chemistry, remains a key challenge to be addressed~\cite{white2023future}.

\section{Challenges}

Although current research on AI for chemistry has achieved very promising results, we should  acknowledge that they are still some way from widespread practical application. Looking ahead, we summarize several key challenges that require further attention for future progress.

\subsection{Data Scarcity}

Since conducting chemical experiments or measurements often entails substantial investments of time and effort, high-quality datasets in this field  remain quite scarce. For example, the largest HTE dataset commonly used for yield prediction contains only a few thousand reaction records~\cite{ahneman2018predicting, perera2018platform}, which is significantly smaller than the training datasets available in fields like computer vision and natural language processing. The limited size of such datasets restrict the ability of models to uncover underlying principles and increases the risks of overfitting, ultimately resulting in poor generalization performance. Active learning can help alleviate this issue by allowing researchers to dynamically supplement the dateset with a small number of meaningful data points, thus improving the model's capability~\cite{settles2009active, ren2021survey, shields2021bayesian, gong2021deepreac+}. In addition, leveraging models from other scenarios through transfer learning~\cite{pan2009survey, zhuang2020comprehensive, shim2022predicting} or pretraining on large but coarse-grained datasets can also be beneficial.

\subsection{Data Bias}

Experimental reaction results reported in the literature often contain various human biases, such as a tendency to record successful cases while overlooking failed experiments (i.e., negative data)~\cite{back2024accelerated}. While   negative data may seem of limited practical value, they are crucial for  training ML models~\cite{jia2019anthropogenic, strieth2022machine}. For example, Moosavi et al.~\cite{moosavi2019capturing} used a robotic synthesis procedure to reconstruct some failed experiments of a meta-organic framework. With the addition of these negative data, they successfully identified the best synthesis condition reported to date. Additionally, reliance on established and available routines can lead to imbalanced data distribution. Moreover, variations in data accuracy and distribution may arise from different measurement instruments and recording methods, leading to biases across data from different sources. These implicit biases can limit the effectiveness of data-driven AI approaches in exploring chemical space. To mitigate such bias, one potential avenue is to establish a community-accepted standard for reporting reaction information, involving both reaction results and other experimental details. For the computer science community, mathematically modeling different types of biases and developing corresponding robust models could be another promising approach. For instance, distributionally robust optimization (DRO)~\cite{Rahimian2019DRO} provides a way to account for data bias by optimizing worst-case performance over a set of possible data distributions. This approach ensures that models remain reliable even when training data distributions deviate from real-world conditions, offering a potential solution to address data heterogeneity in chemical applications..

\subsection{Interpretability}

As discussed above, AI methods have made excellent progress in various applications within chemistry. However, the research  on interpreting the results has received relatively little attention~\cite{bender2022evaluation}. Most current AI methods, especially neural networks, are considered black-box, meaning it is challenging to understand the rationale behind their predictions or decisions. This lack of transparency may lead chemists to question the reliability of results produced by AI models. Furthermore, in certain areas involving safety concerns, such as drug discovery, providing reasonable justification for predictions is of utmost importance. Fortunately, recent studies have started to recognize the importance of interpretability in chemistry and have begun to address this issue. For example, Rodr\'{i}guez-P\'{e}rez and Bajorath~\cite{rodriguez2019interpretation, rodriguez2020interpretation2} used \emph{Shapley Additive exPlanations (SHAP)}~\cite{lundberg2017unified} to interpret compound activity predictions, while Ishida et al.~\cite{ishida2019prediction} employed the \emph{Integrated Gradients} method~\cite{sundararajan2017axiomatic} to highlight reaction-related atoms for retrosynthesis. However, most current studies merely apply general interpretability methods without incorporating chemical principles. We look forward to further developments in this area, which will enhance the practical application of AI methods and, in turn, contribute to advancing the field of chemical science.

\subsection{Generative Large Model}

General-purpose generative large models like GPT have demonstrated outstanding performance across various tasks in natural language processing. However, current evaluations suggest that their performance on domain-specific problems in chemistry is still inadequate~\cite{guo2023can, castro2023large}. As a result, the development of domain-specific generative large models for chemistry is becoming increasingly anticipated. Although several existing models are referred to as chemical large models, they primarily rely on general-purpose LLMs that are fine-tuned on chemistry-related corpora using the next-token prediction as training approach. This training method, which derives distributional dependencies between words from vast textual data, may not be fully suitable for capturing the intricate chemical knowledge required. On one hand, researchers need to consider how to organize chemical data to enhance its compatibility with current general training methods. On the other hand, exploring training approaches specifically tailored to raw chemical data also warrants further attention.

\section{Conclusion}
Overall, this review offers a comprehensive introduction to AI for chemistry from a computational perspective. The article summarizes the current state of AI research in chemistry from three levels: data, representations, and  ML models for various applications, and concludes by highlighting some of the challenges currently faced. We hope this article could provide computer scientists interested in this field with a clear understanding of the current research landscape, and we look forward to the emergence of more groundbreaking work in the future.

\section{Acknowledgement}

This work was supported by the National Natural Science Foundation of China (No. 62272432), the National Key Research and Development Program of China (No. 2021YFA1000900), and the Natural Science Foundation of Anhui Province (No. 2208085MF163). The author also want to thank Prof. Yifeng Wang, Prof. Yunhe Xu, Qiang Zhao, Zheyuan Xu for their helpful comments on this draft.

\bibliography{ref}
\bibliographystyle{plainnat}

\end{document}